# Enhancement of the pinning potential in a single-crystal superconductor FeTe$_{0.65}$Se$_{0.35}$ under the influence of hydrogen sorption

V.P. Timofeev[1], S.I. Bondarenko[1], V.V. Meleshko[1], Yu.A. Savina[1] and A. Wiśniewski[2]

[1] *B.I. Verkin Institute for Low Temperature Physics. National Academy of Sciences of Ukraine, 47 Nauki Ave., Kharkiv, 61103, Ukraine*

[2] *Institute of Physics, Polish Academy of Sciences, Aleja Lotników 32/46, PL-02668 Warsaw, Poland*

E-mail: timofeev@ilt.kharkov.ua; bondarenko@ilt.kharkov.ua

1. Introduction.

The electronic, transport, and magnetic characteristics of promising high-temperature superconductors (HTSC) based on iron-chalcogenides are very sensitive to various doping [1, 2]. For a wide number of compounds, including metallic, cuprate, pnictide and carbon-based materials, an impact of hydrogen causes a noticeable improvement in their superconducting state properties [3-7].

Recently, in order to find ways to increase the current-carrying capacities of HTSC single crystals of FeTe$_{0.65}$Se$_{0.35}$, studies were carried out on the effect of the sorption of gaseous hydrogen on the structure of this type of compounds [8]. Along with structural changes in the crystalline phase, distortions of the crystal lattice of the compound and accumulation of the amorphous state were observed, where the amplitude of the order parameter is locally partially suppressed in the superconducting phase. In these regions, which act as effective pinning centers of magnetic vortices, their averaged potential deepens. The thermal activation energy required for jumps of vortices from the bottom of the potential well increases. This leads to an increase in the maximum attainable current densities in the superconductor [9, 10].

This paper presents the results of contactless SQUID-magnetometric studies of single crystals of the FeTe$_{0.65}$Se$_{0.35}$ compound saturated with hydrogen at temperatures up to 250 °C. The main attention is paid to the description and discussion of the results of changes in the magnetic and superconducting characteristics under the influence of sorption of gaseous hydrogen in a wide range of experimental conditions (namely, the temperatures of the studied single crystal, the conditions of its initial cooling, the values of magnetic fields, etc.).

2. Experimental technique

The experiments were carried out on FeTe$_{0.65}$Se$_{0.35}$ single crystals grown from the melt by the Bridgman method [10]. Hydrogen saturation was performed at several values of temperature, pressure, and residence time of the samples in the gas. The temperature of the samples in the gas varied from room temperature to 250 °C, and the pressure from $10^{-6}$ to 5 atmospheres. The time of interaction of gaseous hydrogen at room temperature in some experiments reached 720 hours.

The magnetic characteristics of a single-crystal iron-chalcogenide compound were studied in the crystal temperature range from 4 K to 300 K. In this case, using an MPMS-5 QD magnetometer, the dependences of the dipole magnetic moment (*m*) of the FeTe$_{0.65}$Se$_{0.35}$ sample on its temperature (*T*), on the degree of doping with hydrogen (which is characterized by the hydrogen gas temperature (°C) during the saturation of the single crystal with $H_2$) and on the magnitude of the applied constant magnetic field (*B*). In the region of the superconducting phase transition (temperature range from 4 K to 20 K), using the measured dependences *m(T)*, the critical temperature of the superconductor ($T_c$) was determined.

The presence of impurities of other, non-basic phases of the compound was estimated from the steepness and monotonicity of the transition *m(T)*. Based on the results of measurements of the isothermal relaxation of the remnant magnetization *m(t)*, the rate of isothermal relaxation of the magnetization (*S*) and the effective value of the pinning potential (*U*) averaged over its volume, a parameter that determines the value of the maximum attainable current of a type II superconductor, were estimated. For this assessment, the Anderson-Kim thermally activated creep model was used [9,10].

3. Research results and their discussion.

It was found that, in the investigated temperature range of 20 °C - 250 °C, with an increase in the sample temperature, the mechanisms of hydrogen sorption and crystal symmetry change. At low temperatures, the tetragonal phase with van der Waals interaction of hydrogen molecules with matrix elements is stable.

Under the influence of the catalytic action of Fe atoms in $FeTe_{0.65}Se_{0.35} + H_2$ solutions at a temperature of about 200 °C, dissociation of hydrogen molecules occurs. This process is accompanied by an increase in the value of mechanical stresses, as well as inhomogeneity of local deformations of interatomic chemical bonds. In the resulting substitutional solutions, due to the large difference in the diameters of interacting particles with an increase in the concentration of hydrogen molecules, an increase in the level of local displacement defects and internal stresses is observed. This leads to the appearance of additional pinning centers during the transition of the sample to the superconducting state [9,10].

In addition, the data obtained [2,3] indicate that, in the crystals under study with an atomic $H^+$ impurity, the relaxation of stresses and local deformations near defects occurs at the phase transition from a tetragonal to an orthorhombic lattice. As could be expected, the transition is accompanied by strong compression of the lattice at which the density of crystals increases by almost 15% [8].

The features of the crystal structure of the compound under study, as well as its modification as a result of hydrogen sorption, lead to noticeable changes in the magnetic and superconducting characteristics of $FeTe_{0.65}Se_{0.35}$. The object of our research was an elongated single-crystal sample of iron-chalcogenide with a maximum size of $6 \times 2.5 \times 0.8$ mm$^3$ (the c axis of the crystal is directed perpendicular to the layers, along the smallest sample size). The test crystal weighs 45 mg. To minimize the effect of the shape of the superconductor on readings of the SQUID magnetometer, the sample was oriented with a large size along the axis of the receiving loops of the flux transformer and, thus, along the axis of the solenoid, which can create the necessary constant magnetic field *H* during the experiment (*H* ∥ *a, b*).

In Fig. 1 shows, as an example, the behavior of the magnetic moment (diamagnetic response) of the compound under study in the region of the superconducting phase transition. The curves of the dependences *m(T)* are shown both for the sample without hydrogen sorption (Fig.1a) and after doping with $H_2$ at a temperature of 150 °C (Fig.1b).

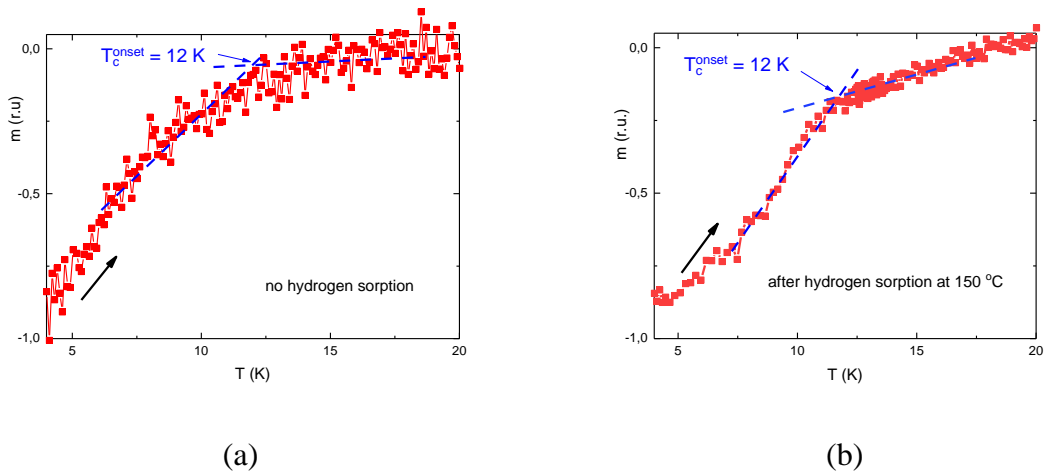

(a)                  (b)

Fig. 1. Dependence of the magnetic moment $m$ (in relative units) of a single crystal on its temperature in the region of the superconducting phase transition: (a) - for the initial sample without $H_2$ sorption; (b) - for a sample doped with hydrogen at 150 °C. The data were taken with an increase in the temperature of the sample after its initial cooling in a small ($B$ = 5 G) constant magnetic field (FC mode). The black bold arrow next to the curves indicates the direction of the crystal temperature change during the measurement of $m(T)$.

The figure shows (in the temperature range from 5 K to 12 K) an increase in the steepness of the superconducting phase transition after hydrogen sorption, which indicates an improvement in the homogeneity of the structural characteristics of the test sample.

An important characteristic of a superconductor, which determines its magnetic and current-carrying abilities, is the ability to hold (pin) magnetic fluxes induced in the surface layer or trapped in the bulk (Abrikosov and Josephson vortices, their bundles of different densities) [9, 10].

To study the dynamics of magnetic fluxes in $FeTe_{0.65}Se_{0.35}$, measurements were carried out in the sample cooling mode in a given uniform magnetic field of the solenoid (FC - field cooling mode). With a decrease in temperature during the experiment and the transition of the sample to the superconducting state, most of the magnetic field is pushed out of the boundaries of the sample, and some, in the form of Abrikosov or Josephson vortices and their bundles, are captured by various defects throughout the volume of the crystal. This technique makes it possible to study the dynamics of trapped magnetic fluxes at $B < B_{c1}$ ($B_{c1}$ is the first critical field of a superconductor).

When the solenoid field is turned off, the remnant magnetization of the sample and its dynamics are determined by the state of the magnetic fluxes trapped by the pinning centers in the bulk of the superconductor. The role of near-surface energy barriers (for example, the Bean-Levingston barrier), which are difficult to control and analyze, in the dynamics of magnetic fluxes under these conditions is minimal.

The thermally activated creep of individual vortices and their bundles leads to redistribution and damping of bulk superconducting currents, the integral dipole moment $m$ begins to decrease with time $t$, and the averaged magnetization $M = m/V$ relaxes (Fig. 2). Here $V$ is the volume of the superconducting sample.

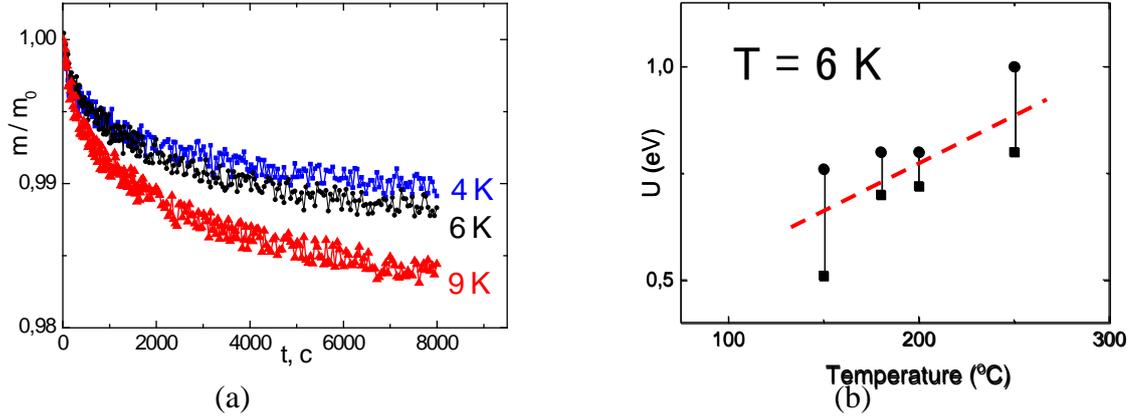

Fig. 2. (a) Curves of isothermal relaxation of the magnetic moment of the sample $m(t)$, normalized to its maximum, initial value $m_0$, for some temperatures of the tested single crystal. The sample was saturated with hydrogen at a temperature of 150 °C. (b) Change in the value of the effective pinning potential $U$ caused by the effect of hydrogen sorption procedures (as an example, the results of calculations of $U$ from measurements of $m(t)$ for a sample temperature $T = 6$ K are given). "Whiskers" (vertical lines) correspond to the scatter of the calculation results.

Having determined from the experimental data in Fig. 2a the normalized rate of isothermal relaxation $S = (1/M_0)\, dM/d\ln(t)$ and using the linear model of thermally activated creep, we can calculate the averaged value of the pinning potential $U = - k_B T/S$. Here: $M_0$ is the initial value of the magnetization, at $t = 0$; $k_B$ is the Boltzmann constant.

From the above Fig. 2 (b) it follows that in the investigated temperature range of the $FeTe_{0.65}Se_{0.35}$ sample under the influence of $H_2$ sorption, the effective pinning potential ($U$) increases with increasing hydrogen temperature. According to the classical equation of thermally activated magnetic flux creep, $J_c/J_{c0} = [1 - (k_B T/U)\ln t]$, where $J_{c0}$ is the critical current density in the absence of thermal fluctuations, which is usually taken as the depairing current. Consequently, in accordance with the collective creep model [9,10] of trapped magnetic fluxes, the critical current density ($J_c$) of the superconductor should increase.

The obtained high values of the thermal activation energy (0.5 eV < U < 1.0 eV) ensure the achievement of an acceptable maximum current $J_c \geq 10^4$ A/cm$^2$ (from the applied point of view) which was previously confirmed for other types of high-temperature superconductors by comparing with transport measurements [10-12]. The results of these studies are in the qualitative measurement of the magnetic characteristics of iron-chalcogenides using an alternating magnetic field [13].

## 4. Conclusion

As a result of experimental studies of the single-crystal iron-chalcogenide compound $FeTe_{0.65}Se_{0.35}$, the effect of structural transitions caused by hydrogen sorption on the magnetic and current-carrying properties of a superconductor has been established. An increase in the volume-averaged effective pinning potential (and the associated critical current density) after the process of hydrogen sorption at temperatures up to 150 °C - 200 °C can be explained by the appearance of additional pinning centers due to the local action of implanted H ions on its crystal structure and electronic states. It was confirmed that hydrogenation is an efficient tool for increasing flux pining properties of superconductors.